# Potentials of ChatGPT for Annotating Vaccine Related Tweets

Md. Rafiul Biswas
College of Science and Engineering
Hamad Bin Khalifa University
Doha, Qatar
0000-0002-5145-1990

Farida Mohsen
College of Science and Engineering
Hamad Bin Khalifa University
Doha, Qatar
0000-0002-0766-4315

Zubair Shah
College of Science and Engineering
Hamad Bin Khalifa University
Doha, Qatar
0000-0001-7389-3274

Wajdi Zaghouani
College of Humanities and Social Sciences
Hamad Bin Khalifa University
Doha, Qatar
0000-0003-1521-5568

*Abstract*—This study evaluates ChatGPT's performance in annotating vaccine-related Arabic tweets by comparing its annotations with human annotations. A dataset of 2,100 tweets representing various factors contributing to vaccine hesitancy was examined. Two domain experts annotated the data, with a third resolving conflicts. ChatGPT was then employed to annotate the same dataset using specific prompts for each factor. The ChatGPT annotations were evaluated through zero-shot, one-shot, and few-shot learning tests, with an average accuracy of 82.14%, 83.85%, and 85.57%, respectively. Precision averaged around 86%, minimizing false positives. The average recall and F1-score ranged from 0.74 to 0.80 and 0.65 to 0.93, respectively. AUC for zero-shot, one-shot, and few-shot learning was 0.79, 0.80, and 0.83. In cases of ambiguity, both human annotators and ChatGPT faced challenges. These findings suggest that ChatGPT holds promise as a tool for annotating vaccine-related tweets.

## I. Introduction

ChatGPT is an advanced language model pre-trained on a massive dataset of diverse domains like human language, including books, articles, and websites. This artificial intelligence (AI) driven model has been improved through reinforcement learning by keeping human feedback in the loop [1]. ChatGPT has revolutionized natural language processing (NLP) techniques by assisting users in generating human-like responses to a wide variety of tasks and queries, ranging from simple questions to more challenging tasks [2]. The potential of ChatGPT has emerged in different sectors to strengthen the capability of humans. Researchers can use ChatGPT as a research assistant in data collection, annotation, analyses, and result interpretation. Researchers can get their desired responses by writing effective prompts to guide ChatGPT. The generative AI model employed in ChatGPT can produce responses that closely resemble human-generated content [3]. ChatGPT has the strength to improve scientific research quality by assisting in writing, accelerating the literature review process, summarizing findings, and enhancing readability [4]. Queries to ChatGPT can speed up the research procedure that requires more effort in designing experimental setup and preparing results [5]. ChatGPT prompts inquiry into their impact on vocabulary and lexical richness; initial comparisons indicate that ChatGPT uses fewer distinct words and lower lexical richness than humans in certain tasks, necessitating further research to understand the broader effects on language evolution across various text types and languages [6]. It extends the use of psychological words by generating a human-like response which was not possible in the traditional vectorized model [7].

The widespread use of AI-generated techniques has disrupted the technology by associating it with new challenges in different domains. For example, ethical concerns remain about facilitating AI's role corresponding to the role of traditional human endeavors in the authorship [8]. Despite various potentials in academic research, ChatGPT is not beyond limitations. Several challenges must be addressed, such as stereotypes, biased responses, misleading information, and ethical considerations [9], [10]. ChatGPT may generate content that could potentially infringe on copyright and intellectual property rights, raising legal concerns. Being a newly developed technology, it still requires more instructions to reach a mature phase. The output of a model might be negatively impacted by the bias in the data and the constraints in the training data [11]. Establishing guidelines is crucial for the integration of ChatGPT in scientific research, ensuring that concerns regarding transparency, reliability, and trustworthiness are thoroughly addressed. Another aspect is that ChatGPT-generated content has fewer distinct words and lacks lexical richness compared to humans [6].

This paper aims to use the ChatGPT as a potential tool for data annotation. Data annotation is the process of giving descriptive labels or metadata so that machine learning models can use it to learn and detect patterns [12]. Data annotation is a time-consuming and challenging process because it requires multiple annotators who work individually to annotate the data [13]. Once the annotation is completed, accuracy and consistency are validated by reviewing a sample of the annotated data. In this paper, vaccine-related Twitter data is annotated with ChatGPT. Twitter data is unstructured and noisy because it does not have a pre-defined format and uses slang, abbreviations, and punctuation [14]. So annotating Twitter data effectively involves several steps in data preprocessing and cleaning. Vaccine-related content are available on Twitter. This is a great source to analyze public sentiment and attitudes toward vaccine hesitancy. Twitter users generate a huge volume of tweets which is quite difficult to annotate by humans. It would be a useful approach if ChatGPT can be employed in annotating vaccine-related tweets. Annotating tweets requires generating appropriate prompts for ChatGPT. This paper shows how accurately ChatGPT is able to annotate vaccine-related Twitter data.

## II. METHOD

We aimed to evaluate the performance of ChatGPT for data annotation tasks. For this, we downloaded vaccine-related Twitter data. Two researchers individually annotated the data and came to the ultimate label by resolving the conflict between them. Then, we employed ChatGPT to annotate the same data and identified the differences between human annotation and ChatGPT annotation. Figure 1 shows a brief of the methodology performed by human annotators and ChatGPT.

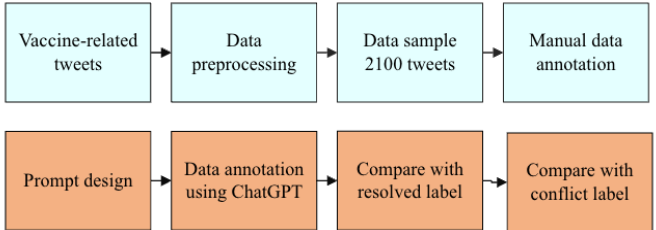

Fig.1. Flow diagram of methodology

### A. Data Collection and Data Cleaning

Vaccine-related Arabic tweets were collected between January 2020 to January 2022. The search terms included vaccine-related keywords in Arabic languages such as "لقاح", "لقاحات", "تطعيم تطعيمات", "لقاح كوفيد-19". Twitter academic research API [15] was used to download the data into the PostgreSQL database. Data were extracted into CSV files using the Python programming language. Then, duplicate entries and retweets were removed. Data were cleaned by removing punctuation, links, mentions, hashtags, and stopwords. Only Arabic tweets were considered for analysis, and so tweets in other languages were removed.

### B. Data Sample

Vaccine-related Twitter data was extensive, and a sample was selected for analysis. Our aim was to identify the stance of tweets that affect vaccine hesitancy. Vaccine hesitancy refers to the reluctance or delay of the vaccination procedure despite the availability and accessibility of vaccination services [16]. The causes that influence vaccine decision-making are called factors. Researchers have developed several models such as epidemiological triads (i.e., external, host, specific), the 5C model, and the health belief model to define the scope and challenges of the vaccine hesitancy [17]. The factors/determinants can be explained in terms of the model. Out of them, we have chosen seven factors such as i) Vaccine safety and efficacy ii) Vaccine acceptance iii) Side effects of vaccines iv) Trust in policymaker role v) Prevention of the risk of infection vi) Vaccine is required to get access in office or travel vii) Influenced by Misinformation/Conspiracy Theory. We applied a cluster sampling technique ( i.e., dividing the population into clusters and then randomly selecting samples from each cluster) [18] to select tweets from each factor. We randomly selected 300 tweets from each factor to analyze the performance of ChatGPT in data annotation. In total, we have 2100 tweets. We divided each factor into training and testing datasets: 20% of the dataset was training and 80% was testing. The training dataset was used to design prompts with high accuracy close to human annotators.

### C. Manual Data Annotation

Authors having domain knowledge (i.e., published several papers on vaccine hesitancy) were employed to annotate data. Second and Third author are native Arabic speakers and so they annotated the Arabic tweet. The first author is not Arabic Speaker and used Google translator to translate it into English and then annotated it. The second author verified the English translation of tweets. For each sub-group, there were three labels: agree, disagree, and neutral. The researcher annotated data individually. The fourth author resolved if there was any conflict in any data. Once the annotation was completed, we validated the accuracy and consistency of the annotations between the two researchers.

### D. Annotate Data Using ChatGPT

We annotated the same data sample (annotated by human annotators) using ChatGPT. We generate a prompt for each factor. Each factor has a different prompt with three labels: agree, disagree, and neutral. High-quality prompt in ChatGPT is important to the language model to generate the desired and relevant responses [19]. Fine-tuning the model on specific tasks or domains can improve ChatGPT performance. We attempted with different prompts on the training dataset and calculated the accuracy of ChatGPT. We selected the best prompt that produced more accurate labels compared to human annotators. Then, we applied this prompt to the test dataset to verify the performance of ChatGPT. The test dataset was never labeled by ChatGPT and was kept separate for Zero-Shot, One-Shot, and Few-Shot learning tests [20].

Zero-shot learning refers to training a model in an unforeseen scenario to classify objects. This approach validates the prior knowledge of ChatGPT to make predictions instead of relying solely on labeled data for training. We give the prompt to ChatGPT and asked them to label it. There were no example tweets that GPT could learn from them. It completely depends on the prompt we have given to it.

One-Shot learning trains a model by providing one example to recognize and classify objects. Besides providing the prompt to ChatGPT, we provided one example tweet from each category of 'agree', 'disagree', and 'Neutral' and asked to label them.

Few-shot learning is similar to one-shot learning, but the model is trained with a limited number of labeled examples of each object. We provided five example tweets along with the prompt and asked ChatGPT to label them.

### E. Prompt Engineering

ChatGPT is highly prompt-dependent. Prompt engineering enhances the capability of ChatGPT by providing human-like-responses [21]. Generating appropriate prompts is essential to utilize the maximum potential of AI language models [22]. We tried different prompts and measured their accuracy. We selected the prompt that produced more accurate results close to annotators. Below is an example of a prompt for zeros-shot learning that we used for the 'vaccine safety and efficacy' factor.

Vaccine safety and efficacy: "Label tweets related to vaccine safety and efficacy as either "Agree," "Disagree," or "Neutral." A tweet expressing belief in the safety and effectiveness of vaccines should be labeled "Agree." A tweet expressing doubt about the safety or effectiveness of vaccines should be labeled "Disagree." A tweet that does not express a clear stance on vaccine safety and efficacy should be labeled "Neutral." Can you label the following tweet?". Similarly, we wrote prompts for one-shot and few-shot learning by adding example tweets for each category of 'Agree', 'Disagree', and

'Neutral'. This procedure was repeated for other factors using different prompts.

## III. RESULTS

This section describes the results obtained from annotators and ChatGPT. It also describes a comparative analysis between human annotator and ChatGPT. This study emphasizes the importance of prompt engineering and provides insights into using ChatGPT effectively in data annotation tasks.

### A. Dataset Description

The dataset contained about ~2.37 million unique Arabic tweets related to vaccines. Tweets that contributed to vaccine hesitancy were considered for this study. We separated tweets according to each factor. The percentages of tweets for each factor are vaccine acceptance (1.4%), vaccine safety and efficacy (4.2%), side effects of vaccines (3.16%), vaccine requirement to get access in office or travel (3.81%), trust in policymaker role in vaccination (5.17%), prevention of the risk of infection (1.34%), influenced by misinformation/conspiracy theory (2.73%). 21.81% of total tweets belonged to the seven factors. The rest of the tweets were unconsidered for analysis due to either falling into other categories or being irrelevant to vaccine topics. After applying the cluster sampling technique, we selected 300 tweets randomly from each factor. Finally, we have 2100 tweets that we labeled both manually and using ChatGPT for analysis.

### B. Conflict for Human Annotators

Two authors annotated the tweets individually, and the third author resolved the conflicts through discussion. It was observed that neutral tweets were conflicting more than others. Also, some tweets were confusing because they carried meanings of both 'agree' and 'disagree' concepts. The annotator misspelled these kinds of tweets. Figure 2 shows the results of the conflict between two annotators for each subgroup. The conflict found between the two annotators are: vaccine acceptance (27%), vaccine safety and efficacy (23%), side effects of vaccines (20%), Vaccine's requirement to get access in the office or travel (26%), trust in policymaker role in vaccination (25%), prevention of the risk of infection (21), and influenced by misinformation/conspiracy (24%).

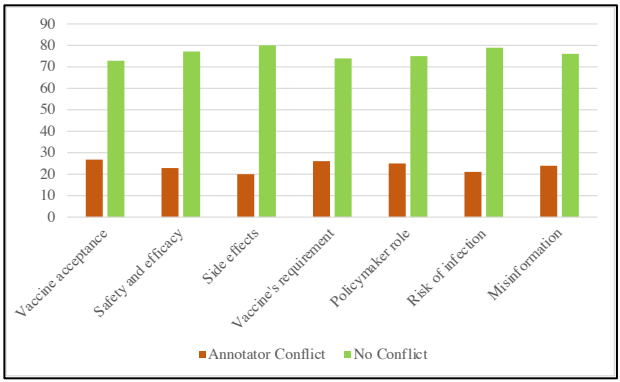

Fig.2. Conflict between two annotators

### C. Evaluation for ChatGPT Annotation

This section shows the performance of ChatGPT on comprehensive data. Every tweet possesses three labels (i.e., agree, disagree, and neutral). The label annotated by ChatGPT was compared with the resolved label (annotated by humans). ChatGPT classifies each tweet into three classes: agree, disagree, and neutral. This experiment was carried out in three tests (e.g., zero-shot, one-shot, and few-shot) to analyze the performance of ChatGPT.

ChatGPT performance model was evaluated by using the confusion matrix. A confusion matrix is a table that evaluates a classification model's performance. It provides a summary of the predictions by combining four different values. The four values are:

True Positive (TP): correctly predicted positive as positive.

True Negative (TN): correctly predicted negative as negative.

False Positive (FP): incorrectly predicted positive as negative.

False Negative (FN): incorrectly predicted negative as positive.

We measured metrics such as accuracy, precision, recall, and F1 score of confusion to evaluate the performance of a ChatGPT classification, as shown below.

$$Accuracy = \frac{TP+TN}{TP+TN+FP+FN} \quad (1)$$

$$Precision = \frac{TP}{TP+FP} \quad (2)$$

$$Recall = \frac{TP}{TP+FN} \quad (3)$$

$$F1-score = \frac{2*(precision*recall)}{(precision+recall)} \quad (4)$$

Accuracy is the fraction of all instances that were correctly classified. Table 1 shows the accuracy of three tests for each factor. It is observed from Table 1 that in every case, the accuracy of a one-shot test is better than a zero-shot test. Similarly, the accuracy of a few-shot test is better than a one-shot test. It demonstrates that the ChatGPT model learns from the data that is used for training. The average accuracy for zero-shot, one-shot, and few-shot learning was 82.14%, 83.85%, and 85.57%, respectively.

Precision predicts positive classes that are actually positive. Table 1 calculates the average precision for zero-shot, one-shot, and few-shot learning as 0.86, 0.865, and 0.86, respectively. The lowest precision was found to be 0.71 for the one-shot test in misinformation. The overall prediction for positive is about 86% accurate (i.e., high precision value), which indicates that the ChatGPT is good at avoiding false positives in data annotation tasks.

Recall measures the strength of a model to identify all of the positive cases. A high recall is observed at 0.97 for zero-shot in risk prevention, and a low recall is observed at 0.56 zero-shot in vaccine's requirement to get access. A high recall value indicates that the model is good at identifying true positives, while a low recall value implies that the model is missing some positive cases. The average recall for zero-shot, one-shot, and few-shot is 0.74, 0.77, and 0.80.

F1-score, a widely used evaluation metric in binary classification, combines precision and recall into a single measure. A higher F1-score signifies a better balance between precision and recall, suggesting that the model has performed well in correctly identifying positive classes and capturing all positive ones. The highest and lowest F1-score are observed at 0.93 and 0.65 in zero-shot learning in preventing the risk of infection and one-shot learning influenced by misinformation, respectively.

Area Under Curve (AUC) evaluates the performance of a model in binary classification across different threshold settings. Basically, it points to the area under the receiver operating characteristic curve (ROC curve). A higher AUC implies better discriminative power of classification. The average AUC for zero-shot, one-shot, and few-shot learning is 0.79, 0.80, and 0.83 shown in Table 1. The abbreviations in Table 1 and Table 2 refer to ZS-Zero Shot, OS-One Shot, and FS-Few Shot.

TABLE I. ChatGPT Performance on Overall Data

| Factor | Test | Accuracy | Precision | Recall | F1-score | AUC |
|---|---|---|---|---|---|---|
| Vaccine acceptance | ZS | 0.89 | 0.90 | 0.89 | 0.89 | 0.90 |
|  | OS | 0.92 | 0.89 | 0.89 | 0.89 | 0.91 |
|  | FS | 0.93 | 0.90 | 0.89 | 0.89 | 0.93 |
| Vaccine safety and efficacy | ZS | 0.87 | 0.86 | 0.86 | 0.86 | 0.85 |
|  | OS | 0.88 | 0.87 | 0.87 | 0.87 | 0.90 |
|  | FS | 0.89 | 0.89 | 0.88 | 0.88 | 0.91 |
| Side effects of vaccines | ZS | 0.79 | 0.89 | 0.76 | 0.82 | 0.82 |
|  | OS | 0.79 | 0.86 | 0.64 | 0.69 | 0.86 |
|  | FS | 0.85 | 0.86 | 0.66 | 0.68 | 0.86 |
| Vaccine's requirement to get access | ZS | 0.83 | 0.86 | 0.56 | 0.68 | 0.82 |
|  | OS | 0.84 | 0.83 | 0.65 | 0.73 | 0.72 |
|  | FS | 0.85 | 0.78 | 0.89 | 0.67 | 0.86 |
| Trust in policymaker's role in vaccination | ZS | 0.76 | 0.83 | 0.62 | 0.71 | 0.67 |
|  | OS | 0.79 | 0.83 | 0.62 | 0.71 | 0.72 |
|  | FS | 0.81 | 0.97 | 0.70 | 0.81 | 0.74 |
| Prevention of the risk of infection | ZS | 0.89 | 0.99 | 0.89 | 0.93 | 0.79 |
|  | OS | 0.89 | 0.80 | 0.97 | 0.89 | 0.80 |
|  | FS | 0.89 | 0.88 | 0.88 | 0.87 | 0.81 |
| Conspiracy theory | ZS | 0.72 | 0.73 | 0.66 | 0.65 | 0.69 |
|  | OS | 0.76 | 0.71 | 0.79 | 0.81 | 0.73 |
|  | FS | 0.77 | 0.74 | 0.76 | 0.89 | 0.76 |

TABLE II. ChatGPT Performance Evaluation on Conflict Data

| Factor | Test | Accuracy | Precision | Recall | F1-score | AUC |
|---|---|---|---|---|---|---|
| Vaccine acceptance | AC | 0.88 | 0.88 | 0.87 | 0.87 | 0.84 |
|  | NC | 0.89 | 0.89 | 0.88 | 0.89 | 0.91 |
| Vaccine safety and efficacy | AC | 0.47 | 0.48 | 0.50 | 0.56 | 0.57 |
|  | NC | 0.96 | 0.90 | 0.95 | 0.92 | 0.96 |
| Side effects of vaccines | AC | 0.69 | 0.67 | 0.67 | 0.66 | 0.62 |
|  | NC | 0.84 | 0.79 | 0.74 | 0.81 | 0.82 |
| Vaccine's requirement to get access | AC | 0.56 | 0.57 | 0.52 | 0.57 | 0.50 |
|  | NC | 0.84 | 0.83 | 0.65 | 0.73 | 0.72 |
| Trust in policymaker's role in vaccination | AC | 0.56 | 0.58 | 0.57 | 0.53 | 0.61 |
|  | NC | 0.76 | 0.72 | 0.73 | 0.72 | 0.84 |
| Prevention of the risk of infection | AC | 0.79 | 0.71 | 0.76 | 0.75 | 0.68 |
|  | AC | 0.91 | 0.79 | 0.96 | 0.83 | 0.81 |
| Conspiracy theory | AC | 0.73 | 0.67 | 0.76 | 0.75 | 0.66 |
|  | NC | 0.78 | 0.84 | 0.77 | 0.81 | 0.8 |

### D. Evaluation of ChatGPT in a Conflict Scenario between Annotators

We wanted to observe how ChatGPT works when there is a conflict of annotation between two annotators. We grouped each factor into two categories: i) there is a conflict between two annotators, and ii) there is no conflict between annotators. We performed zero-shot learning for each category of tweets and analyzed the accuracy, precision, recall, F1-score, and AUC. The results are quite interesting: ChatGPT performance degrades when there is a conflict between human annotators. Because the tweets were ambiguous, human annotators also struggled to label them. When there is no conflict between human annotators, ChatGPT also performs better. It happened because the tweets were easy to understand, so there was no conflict between annotators. Similarly, it was also easy for ChatGPT to understand the tweets and label them correctly. Table 2. shows the comparative results of each factor by ChatGPT. There is a significant difference in accuracy between the two categories. For the annotator conflict in the vaccine safety and efficacy factor, accuracy was found to be 0.47. In comparison, accuracy was measured at 0.96 for the no-conflict scenario. Whereas AUC was calculated at 0.57 and 0.96 for conflict and no-conflict scenarios. Similarly, noticeable differences were found for precision, recall, and F1-score between conflict and no-conflict. The abbreviation in Table 2 refers to AC- Annotator Conflict and NC-No conflict.

### E. ROC Curve Analysis on Overall Data

We measured ChatGPT classification performance on overall datasets of seven factors (2100 tweets) by analyzing Receiver Operating Characteristic (ROC) curve. ROC curve graphically represents classification models by plotting the true positive rate (TPR) against the false positive rate (FPR) at various classification thresholds.

$$TPR = \frac{TP}{TP+FN} \qquad (5)$$
$$FPR = \frac{FP}{FP+TN} \qquad (6)$$

It explains how much a model is capable in binary classification task. The higher AUC score, the higher capability to predict classification. Our datasets has three class label: agree, disagree and neutral. For this multiclass classification, we followed 'one-vs-rest' that compares each class against all the other classes. We applied logistic regression for prediction of the ROC curve.

Figure 3 shows a ROC curve for three types of classification: agree, disagree, and neutral for Zero-Shot learning. The model has AUC score of 0.83, 0.79 and 0.85 for agree, disagree and neutral class respectively. Figure 4 shows ROC curve for One-Shot learning where the AUC score is measured as 0.86 for all three classification: agree, disagree and neutral. Figure 5 shows the ROC curve for Few-Shot learning where the AUC score is measured as 0.86, 0.90 and 0.87 for agree, disagree and neutral class respectively. It indicates that ChatGPT, a sophisticated language model, can be used in the context of data annotation tasks.

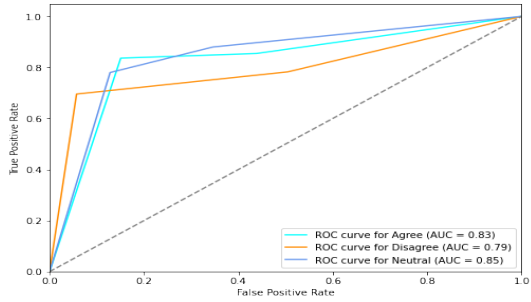

Fig.3. ROC curve for One-vs-Rest classification (Zero-Shot)

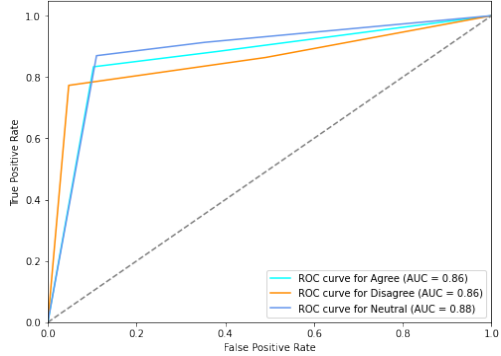

Fig.4. ROC curve for One-vs-Rest classification (One-Shot)

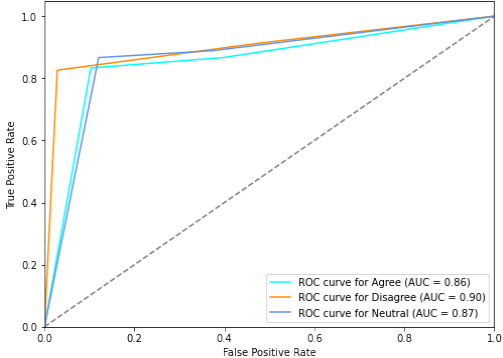

Fig.5. ROC curve for One-vs-Rest classification (Few-Shot)

## IV. Discussion

Data annotation is time-consuming and costly and often requires expert efforts [23]. Researchers have developed machine-learning and deep-learning techniques to annotate data for specific fields. However, the inconsistency among the annotators may significantly reduce the performance of the model [24]. This study performed a comparative analysis between humans and ChatGPT for data annotation tasks to find an alternative way. The results of the study demonstrate the potential of ChatGPT as a tool for annotating vaccine-related tweets. The results suggest that ChatGPT can be used to identify the stance of vaccine-hesitant tweets in real-time. It will help to monitor public attitudes towards vaccines in social media for new tweets. ChatGPT would be able to annotate new categories of tweets if proper prompts were designed by human experts. As ChatGPT learns from humans in the loop [3], any tweets that are identified as being vaccine-hesitant could then be flagged for further review by human experts. The results also suggest that ChatGPT could be used to help develop vaccine education and outreach materials. By understanding the reasons why people are hesitant to get vaccinated, ChatGPT could be used to create materials that are tailored to address these concerns.

The performance of ChatGPT in data annotation was evaluated using different tests, including zero-shot, one-shot, and few-shot learning. The accuracy, precision, recall, F1-score, and AUC were measured to evaluate the performance of ChatGPT in classifying tweets into agree, disagree, or neutral categories. Overall, ChatGPT showed promising results in annotating vaccine-related tweets. The average accuracy for zero-shot, one-shot, and few-shot learning was 82.14%, 83.85%, and 85.57%, respectively. This indicates that ChatGPT can effectively classify tweets with a reasonable level of accuracy. The precision values were high, ranging from 0.86 to 0.89, indicating that ChatGPT is good at avoiding false positives in data annotation. The recall values ranged from 0.74 to 0.80, showing that ChatGPT is capable of identifying true positives, although there were some variations across different factors. The F1-score ranged from 0.65 to 0.93, indicating a good balance between precision and recall for most factors. The AUC values ranged from 0.79 to 0.83, suggesting a good discriminative power of ChatGPT in classifying tweets.

It is worth noting that the performance of ChatGPT varied across different factors. Some factors, such as vaccine acceptance, safety, and efficacy, showed relatively higher accuracy and F1 scores, indicating that ChatGPT performed well in classifying tweets related to these topics. On the other hand, factors like side effects of vaccines and vaccine's requirement to get access showed slightly lower accuracy and F1-scores, suggesting that these topics might be more challenging for ChatGPT to classify accurately. The performance of ChatGPT in trust in the policymaker's role in the vaccination factor was relatively lower, indicating that this topic might require further improvements in the annotation process.

Prompt engineering played a crucial role in enhancing the performance of ChatGPT. By providing clear and specific prompts for each factor, ChatGPT was able to generate more accurate annotations. The choice of prompts significantly influenced the performance of ChatGPT, highlighting the importance of prompt design in maximizing the potential of AI language models. While ChatGPT showed promising results, it is important to acknowledge the limitations and challenges associated with using AI language models for data annotation. One limitation is the reliance on the quality of the training data. If the training data contains biases or inaccuracies, it can affect the performance of ChatGPT in annotation tasks. Therefore, it is crucial to ensure the quality and representativeness of the training data.

Ethical considerations should also be taken into account when using AI language models for data annotation. It is important to address issues such as biases, stereotypes, and misleading information that can be propagated through the model's responses. Guidelines and best practices need to be developed to ensure the transparency, reliability, and trustworthiness of the annotations generated by ChatGPT.

## V. Limitations and Future Works

Twitter data is unstructured and incomplete and sometimes does not belong to any meaningful sentence. Annotating this ambiguous data was challenging and may be misinterpreted by the human annotator. Thus, it would also generate the wrong output for ChatGPT. Usually, this scenario happens when a tweet is posted as a reply to another tweet.

Therefore, it needs to dig deep into the original tweet to find the context of the tweet.

Our data was tested with GPT-3.5, and it has limited knowledge about the world after 2021. Therefore, it is expected that some wrong output while annotation. ChatGPT has limitations on the prompt size, so the prompt should be designed in a concise manner. GPT-4 can be used, but it has a limitation on the number of prompts per hour. Also, ChatGPT occasionally generates the wrong output. As a large language model, the ChatGPT is always learning and expected to generate better outcomes in the future. Allowing developers to finetune GPT-3.5 and GPT-4 models will generate better results for the data annotation task.

## VI. Conclusion

This study demonstrated the potential of using ChatGPT as a tool for annotating vaccine-related tweets. The results showed that ChatGPT could effectively classify vaccine-hesitant tweets into agree, disagree, or neutral categories with reasonable accuracy, precision, recall, and F1-score. This information can be used to target vaccine education and outreach efforts to those who are most likely to be hesitant. Prompt engineering and careful consideration of ethical concerns is essential for maximizing the performance and reliability of ChatGPT in data annotation tasks. Further research and development in this area can contribute to the advancement of AI-driven annotation methods and support research in various domains, including public health and social sciences.

## Acknowledgment

This publication was partially funded by NPRP grants 14C-0916-210015 and NPRP13S-0206-200281 from the Qatar National Research Fund (a member of Qatar Foundation). The findings herein are solely the responsibility of the authors.

## References

[1] C. Celemin, G. Maeda, J. Ruiz-del-Solar, J. Peters, and J. Kober, "Reinforcement learning of motor skills using Policy Search and human corrective advice," Int. J. Rob. Res., vol. 38, no. 14, pp. 1560–1580, 2019, doi: 10.1177/0278364919871998.

[2] S. S. Gill and R. Kaur, "ChatGPT: Vision and challenges," Internet Things Cyber-Physical Syst., vol. 3, pp. 262–271, 2023, doi: 10.1016/j.iotcps.2023.05.004.

[3] Y. K. Dwivedi et al., "'So what if ChatGPT wrote it?' Multidisciplinary perspectives on opportunities, challenges and implications of generative conversational AI for research, practice and policy," Int. J. Inf. Manage., vol. 71, no. March, 2023, doi: 10.1016/j.ijinfomgt.2023.102642.

[4] M. Sallam, "Practice : Systematic Review on the Promising Perspectives and Valid Concerns," no. Ml, 2023.

[5] Z. Liu et al., "DeID-GPT: Zero-shot Medical Text De-Identification by GPT-4," no. Llm, 2023, [Online]. Available: http://arxiv.org/abs/2303.11032.

[6] P. Reviriego, J. Conde, E. Merino-Gómez, G. Martínez, and J. A. Hernández, "Playing with Words: Comparing the Vocabulary and Lexical Richness of ChatGPT and Humans," pp. 1–6, 2023, [Online]. Available: https://arxiv.org/abs/2308.07462.

[7] S. M. Seals and V. L. Shalin, "Long-form analogies generated by chatGPT lack human-like psycholinguistic properties," 2023, [Online]. Available: http://arxiv.org/abs/2306.04537.

[8] H. Else, "Abstracts written by ChatGPT fool scientists," Nature, vol. 613, no. 7944, p. 423, 2023, doi: 10.1038/d41586-023-00056-7.

[9] A. Haleem, M. Javaid, and R. P. Singh, "An era of ChatGPT as a significant futuristic support tool: A study on features, abilities, and challenges," BenchCouncil Trans. Benchmarks, Stand. Eval., vol. 2, no. 4, p. 100089, 2022, doi: 10.1016/j.tbench.2023.100089.

[10] M. Farrokhnia, S. K. Banihashem, O. Noroozi, and A. Wals, "A SWOT analysis of ChatGPT: Implications for educational practice and research," Innov. Educ. Teach. Int., vol. 00, no. 00, pp. 1–15, 2023, doi: 10.1080/14703297.2023.2195846.

[11] S. Mitrović, D. Andreoletti, and O. Ayoub, "ChatGPT or Human? Detect and Explain. Explaining Decisions of Machine Learning Model for Detecting Short ChatGPT-generated Text," pp. 1–11, 2023, [Online]. Available: http://arxiv.org/abs/2301.13852.

[12] M. J. Willemink et al., "Preparing medical imaging data for machine learning," Radiology, vol. 295, no. 1, pp. 4–15, 2020, doi: 10.1148/radiol.2020192224.

[13] A. Chortaras et al., "WITH: Human-Computer Collaboration for Data Annotation and Enrichment," Web Conf. 2018 - Companion World Wide Web Conf. WWW 2018, pp. 1117–1125, 2018, doi: 10.1145/3184558.3191544.

[14] S. Neelakandan and D. Paulraj, "A gradient boosted decision tree-based sentiment classification of twitter data," Int. J. Wavelets, Multiresolution Inf. Process., vol. 18, no. 4, 2020, doi: 10.1142/S0219691320500277.

[15] Twitter, "Twitter API for Academic Research | Products | Twitter Developer Platform." https://developer.twitter.com/en/products/twitter-api/academic-research (accessed May 30, 2021).

[16] N. E. MacDonald et al., "Vaccine hesitancy: Definition, scope and determinants," Vaccine, vol. 33, no. 34, pp. 4161–4164, 2015, doi: 10.1016/j.vaccine.2015.04.036.

[17] D. Kumar, R. Chandra, M. Mathur, S. Samdariya, and N. Kapoor, "Vaccine hesitancy: Understanding better to address better," Isr. J. Health Policy Res., vol. 5, no. 1, pp. 1–8, 2016, doi: 10.1186/s13584-016-0062-y.

[18] G. Sharma, "Pros and cons of different sampling techniques. International journal of applied research," Int. J. Appl. Res., vol. 3, no. 7, pp. 749–752, 2017, [Online]. Available: www.allresearchjournal.com.

[19] J. White et al., "A Prompt Pattern Catalog to Enhance Prompt Engineering with ChatGPT," 2023, [Online]. Available: http://arxiv.org/abs/2302.11382.

[20] F. Sung, Y. Yang, L. Zhang, T. Xiang, P. H. S. Torr, and T. M. Hospedales, "Learning to Compare: Relation Network for Few-Shot Learning," Proc. IEEE Comput. Soc. Conf. Comput. Vis. Pattern Recognit., pp. 1199–1208, 2018, doi: 10.1109/CVPR.2018.00131.

[21] C. E. Short and J. C. Short, "The artificially intelligent entrepreneur: ChatGPT, prompt engineering, and entrepreneurial rhetoric creation," J. Bus. Ventur. Insights, vol. 19, no. March, p. e00388, 2023, doi: 10.1016/j.jbvi.2023.e00388.

[22] L. S. Lo, "The CLEAR path: A framework for enhancing information literacy through prompt engineering," J. Acad. Librariansh., vol. 49, no. 4, p. 102720, 2023, doi: 10.1016/j.acalib.2023.102720.

[23] C. B. Rasmussen, K. Kirk, and T. B. Moeslund, "The Challenge of Data Annotation in Deep Learning—A Case Study on Whole Plant Corn Silage," Sensors, vol. 22, no. 4, pp. 1–19, 2022, doi: 10.3390/s22041596.

[24] C. Schreiner, K. Torkkola, M. Gardner, and K. Zhang, "Using machine learning techniques to reduce data annotation time," Proc. Hum. Factors Ergon. Soc., pp. 2438–2442, 2006, doi: 10.1177/154193120605002219.